\documentclass[runningheads]{llncs}

\usepackage{graphicx}
\usepackage{ucs}
\usepackage[utf8]{inputenc}
\usepackage[american]{babel}
\usepackage{fontenc}

\usepackage{amsmath,amsfonts,amssymb}

\usepackage{enumerate}

\usepackage{hyperref}

\usepackage[linesnumbered,ruled,noend]{algorithm2e}
\newcommand{\nat}{\ensuremath{\mathbb{N}}}
\newcommand{\expected}[2]{\ensuremath{\mathbb{E}_{#1}\left[#2\right]}}
\newcommand{\prob}[2]{\ensuremath{\mathbb{P}_{#1}\left[#2\right]}}
\newcommand{\set}[1]{\ensuremath{\left\{ #1 \right\}}}

\newcommand{\Opt}{\text{\rmfamily\scshape Opt}}
\newcommand{\ALG}{\text{\rmfamily\scshape Alg}}

\newcommand{\eee}{\ensuremath{\mathrm{e}}}
\DeclareMathOperator{\Bin}{Bin}

\usepackage{mathtools}
\usepackage{tikz}
\providecommand{\raolong}{Reading Articles Online}
\providecommand{\rao}{RAO}
\providecommand{\kph}{KPH}

\newtheorem{assumption}[theorem]{Assumption}

\begin{document}

\title{Reading Articles Online\texorpdfstring{\thanks{This paper has been accepted at COCOA 2020.
The final authenticated publication is available online at \url{https://doi.org/10.1007/978-3-030-64843-5_43}}}{}}


\author{Andreas Karrenbauer\inst{1} \and
Elizaveta Kovalevskaya\inst{1,2}}
\authorrunning{A. Karrenbauer and E. Kovalevskaya}

\institute{Max Planck Institute for Informatics, Saarland Informatics Campus, Germany \\ \email{firstname.lastname@mpi-inf.mpg.de} 
\and Goethe University Frankfurt, Germany \\ \email{lisa@ae.cs.uni-frankfurt.de}
}

\maketitle              % typeset the header of the contribution

%\begin{center}
%This paper has been accepted at COCOA 2020.\\
%The final authenticated publication is available online at \url{https://doi.org/10.1007/978-3-030-64843-5_43}
%\end{center}

\begin{abstract}
We study the online problem of reading articles that are listed in an aggregated form in a dynamic stream, e.g., in news feeds, as abbreviated social media posts, or in the daily update of new articles on arXiv. In such a context, the brief information on an article in the listing only hints at its content. We consider readers who want to maximize their information gain within a limited time budget, hence either discarding an article right away based on the hint or accessing it for reading. The reader can decide at any point whether to continue with the current article or skip the remaining part irrevocably. In this regard, \raolong{}, \rao{}, does differ substantially from the Online Knapsack Problem, but also has its similarities. Under mild assumptions, we show that any $\alpha$-competitive algorithm for the Online Knapsack Problem in the random order model can be used as a black box to obtain an $(\eee + \alpha)C$-competitive algorithm for \rao{}, where $C$ measures the accuracy of the hints with respect to the information profiles of the articles. Specifically, with the current best algorithm for Online Knapsack, which is $6.65<2.45\eee$-competitive, we obtain an upper bound of $3.45\eee C$ on the competitive ratio of \rao{}. Furthermore, we study a natural algorithm that decides whether or not to read an article based on a single threshold value, which can serve as a model of human readers. We show that this algorithmic technique is $O(C)$-competitive. Hence, our algorithms are constant-competitive whenever the accuracy $C$ is a constant.
\end{abstract}

\section{Introduction}

There are many news aggregators available on the Internet these days. However, it is impossible to read all news items within a reasonable time budget. Hence, millions of people face the problem of selecting the most interesting articles out of a news stream. They typically browse a list of news items and make a selection by clicking into an article based on brief information that is quickly gathered, e.g., headline, photo, short abstract. They then read an article as long as it is found interesting enough to stick to it, i.e., the information gain is still sufficiently high compared to what is expected from the remaining items on the list. If not, then the reader goes back to browsing the list, and the previous article is discarded -- often irrevocably due to the sheer amount of available items and a limited time budget.
This problem is inspired by research in Human-Computer Interaction~\cite{freire2019foraging}.

In this paper, we address this problem from a theoretical point of view. To this end, we formally model the \raolong{} Problem, \rao{}, show lower and upper bounds on its competitive ratio, and analyze a natural threshold algorithm, which can serve as a model of a human reader.  

There are obvious parallels to the famous Secretary Problem, i.e., if we could only afford to read one article, we had a similar problem that we had to make an irrevocable decision without knowing the remaining options. However, in the classical Secretary Problem, it is assumed that we obtain a truthful valuation of each candidate upon arrival. But in our setting, we only get a hint at the content, e.g., by reading the headline, which might be a click bait. However, if there is still time left from our budget after discovering the click bait, we can dismiss that article and start browsing again, which makes the problem fundamentally more general. Moreover, a typical time budget allows for reading more than one article, or at least a bit of several articles, perhaps of different lengths. Thus, our problem is also related to Online Knapsack but with uncertainty about the true values of the items. Nevertheless, we assume that the reader obtains a hint of the information content before selecting and starting to read the actual article. This is justified because such a hint can be acquired from the headline or a teaser photo, which is negligible compared to the time it takes to read an entire article. In contrast, the actual information gain is only realized while reading and only to the extent of the portion that has already been read. For the sake of simplicity, one can assume a sequential reading strategy where the articles are read word for word and the information gain might fluctuate strongly, especially in languages like German where a predicate/verb can appear at the end of a long clause. However, in contrast to spatial information profiles, one can also consider temporal information profiles where the information gain depends on the reading strategy, e.g., cross reading. It is clear that the quality of the hint in relation to the actual information content of the article is a decisive factor for the design and analysis of corresponding online algorithms. We argue formally that the hint should be an upper bound on the information rate, i.e., the information gain per time unit. Moreover, we confirm that the hint should not be too far off the average information rate to achieve decent results compared to the offline optimum where all articles with the corresponding information profiles are known in advance. In this paper, we assume that the length of an article, i.e., the time it takes to read it to the end, is revealed together with the hint. This is a mild assumption because this attribute can be retrieved quickly by taking a quick glance at the article, e.g., at the number of pages or at the size of the scroll bar.

\subsection{Related Work}
To the best of our knowledge, the problem of \rao{} has not been studied in our suggested setting yet.
The closest related problem known is the Online Knapsack Problem \cite{AlbersKnapsackGAP,KnapsackSecretaryProblem,KesselheimPrimalDual} in which an algorithm has to fill a knapsack with restricted capacity
while trying to maximize the sum of the items' values.
Since the input is not known at the beginning and an item is not selectable later than its arrival, optimal algorithms do not exist.
In the adversarial model where an adversary chooses the order of the items to arrive,
it has been shown in \cite{StochasticOnline} that the competitive ratio is unbounded.
Therefore, we consider the random order model where a permutation of the input is chosen uniformly at random.

A special case of the Online Knapsack Problem is the well-studied Secretary Problem, solved by \cite{Dynkin} among others.
The goal is to choose the best secretary of a sequence without knowing what values the remaining candidates will have. The presented $\eee$-competitive algorithm is optimal for this problem.
The $k$-Secretary Problem aims to hire at most $k\geq 1$ secretaries while maximizing the sum of their values. In
\cite{MCSecretaryAlgorithm}, an algorithm with a competitive ratio of 
$1/(1-5/\sqrt{k})$, for large enough $k$,
with a matching lower bound of $\Omega(1/(1-1/\sqrt{k}))$ is presented.
Furthermore, \cite{KnapsackSecretaryProblem} contains an algorithm that is $\eee$-competitive for any $k$.
Some progress for the case of small $k$ was made in~\cite{albers_et_al:LIPIcs:2019:11514}.
The Knapsack Secretary Problem introduced by \cite{KnapsackSecretaryProblem} is equivalent to the 
Online Knapsack Problem in the random order model.
They present a $10\eee$-competitive algorithm. An $8.06$-competitive algorithm ($8.06<2.97\eee$) is shown in~
\cite{KesselheimPrimalDual} for the Generalized Assignment Problem, which is the Online Knapsack Problem generalized to a setting with multiple knapsacks with different capacities.
The current best algorithm from \cite{AlbersKnapsackGAP} achieves a competitive ratio of $6.65< 2.45\eee$.

There have been different approaches to studying the Knapsack Problem with uncertainty besides the random order model.
One approach is the Stochastic Knapsack Problem where values or weights are drawn from a known distribution.
This problem has been studied in both online \cite{StochasticOnline} and offline \cite{StochasticOffline} settings.
In \cite{TheRobustKnapsackProblemWithQueries}, an offline setting with unknown weights is considered:
algorithms are allowed to query a fixed number of items to find their exact weight.

A model with resource augmentation is considered for the fractional version of the Online Knapsack Problem in \cite{fractionalKnapsack}.
There, the knapsack of the online algorithm has $1\leq R\leq 2$ times more capacity than the knapsack of the offline optimum.
Moreover, they allow items to be rejected after being accepted.
In our model, this would mean that the reader gets time returned after having already read an article.
Thus, their algorithms are not applicable on \rao{}.

\subsection{Our Contribution}

We introduce \rao{} and prove lower and upper bounds on competitive ratios under various assumptions. We present relations to the Online Knapsack problem and show how ideas from that area can be adapted to \rao{}. Our emphasis lies on the initiation of the study of this problem by the theory community. 

We first show lower bounds that grow with the number of articles unless restrictions apply that forbid the corresponding bad instances. That is, whenever information rates may be arbitrarily larger than the hint of the corresponding article, 
any algorithm underestimates the possible information gain of a good article.
Hence, the reader must adjust the hints such that they upper bound the information rate to allow for bounded competitive ratios.
While we may assume w.l.o.g.~for the Online Knapsack Problem that no item is larger than the capacity since an optimal solution cannot contain such items, we show that \rao{} without this or similar restrictions suffers from a lower bound of $\Omega(n)$, i.e., any algorithm is arbitrarily bad in the setting where articles are longer than the time budget.  
Moreover, we prove that the accuracy of the hints provides a further lower bound for the competitive ratio. 
We measure this accuracy as the maximum ratio $C$ of hints and respective average information rates.
Hence, a constant-competitive upper bound for \rao{} is only possible when $C$ is bounded.

Under these restrictions, we
present the first constant-com\-pe\-ti\-ti\-ve algorithm for \rao{}. To this end, we introduce a framework for wrapping any black box algorithm for the Online Knapsack Problem to work for \rao{}. 
Given an $\alpha$-competitive algorithm for the Online Knapsack Problem as a black box, 
we obtain a $(\eee+\alpha)C$-competitive algorithm for \rao{}.
This algorithm is $3.45\eee C$-competitive when using the current best algorithm for the Online Knapsack Problem from \cite{AlbersKnapsackGAP}.
This is the current best upper bound that we can show for \rao{}, which is constant provided that the hints admit a constant accuracy.

However, the algorithm generated by the framework above inherits its  complexity from the black box algorithm for the Online Knapsack Problem, which may yield good competitive ratios from a theoretical point of view but might be too complex to serve as a strategy for a human reader. Nevertheless, the existence of constant-competitive ratios (modulo accuracy of the hints) motivates us to strive for simple $O(C)$-algorithms. To this end, we investigate an algorithm that bases its decisions on a single threshold. The Threshold Algorithm can be seen as a formalization of human behavior. While reading, humans decide intuitively whether an article is interesting or not. This intuition is modeled by the single threshold. In case of diminishing information gain, we show that this simplistic approach suffices to obtain an upper bound of $246 C< 90.5 \eee C$ on the competitive ratio with the current analysis, which might leave room for improvement but nevertheless achieves a constant competitive ratio. 
Diminishing information gain means non-increasing information rates, a reasonable assumption particularly in the light of efficient reading strategies, where an article is not read word for word. In such a context, one would consider a temporal information profile that relates the information gain to the reading time. When smoothed to coarse time scale, the information rates can be considered non-increasing and lead to a saturation of the total information obtained from an article over time. 

\section{Preliminaries}

\begin{definition}[\raolong{} (\rao{})] \label{def:RA}
There are $n$ articles that are revealed one by one in a round-wise fashion. The reader has a time budget $T\in\nat_{>0}$ for reading.
In round $i$, the reader sees article $i$ with its hint $h_i \in\nat_{>0}$ and time length $t_i\in\nat_{>0}$.
The actual information rate $c_i:[t_i]\to [h_i]$ is an unknown function.
The reader has to decide whether to start reading the article or to skip to the next article.
After reading time step $j\leq t_i$, the reader obtains $c_i(j)$ information units and can decide to read the next time step of the article or to discard it irrevocably.
After discarding or finishing the current article, the next round begins.
The objective is to maximize $\sum_{i\in[n]} \sum_{j=1}^{\tau_i} c_i(j)$ where $0\le \tau_i\le t_i$ is the number of time steps read by the algorithm and $\sum_{i\in[n]} \tau_i\leq T$.
\end{definition}

For the sake of simplicity, we have chosen a discrete formulation in Def.~\ref{def:RA}, which is justified by considering words or even characters as atomic information units. Since such tiny units might be too fine-grained compared to the length of the articles, we can also extend this formulation with a slight abuse of notation and allow that the $\tau_i$ are fractional, i.e., $\sum_{j=1}^{\tau_i} c_i(j) = \sum_{j=1}^{\lfloor\tau_i\rfloor} c_i(j) + c_i(\lceil \tau_i\rceil)\cdot \{\tau_i\}$, where $\{\tau_i\}$ denotes its fractional part. However, one could also consider a continuous formulation using integrals, i.e., the objective becomes $\sum_{i\in[n]} \int_{0}^{\tau_i} c_i(t) dt$. The lower bounds presented in this section hold for these models as well.

We use the random order model
where input order corresponds to a permutation $\pi$ chosen uniformly at random.

\begin{definition}[Competitive Ratio]
We say that an algorithm $\ALG$ is $\alpha$-competitive,
if, for any instance $I$, the expected value of $\ALG$ on instance $I$, with respect to permutation of the input and random choices of $\ALG$, is at least $1/\alpha$ 
of the optimal offline value $\Opt(I)$, i.e., $\expected{}{\ALG(I)} \geq \frac{1}{\alpha}\cdot \Opt(I).$
\end{definition}

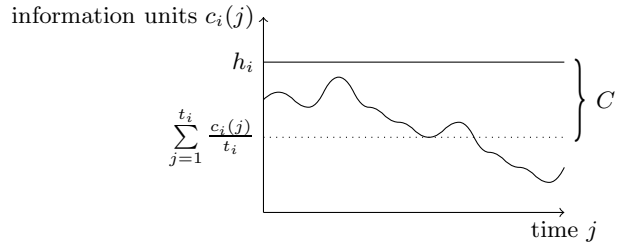
\begin{figure}[tp] 
    \centering
\begin{tikzpicture}[scale=0.20]
    \draw [->] (0,0) -- (0,13) node[align=center, left] {information units $c_i(j)$};
    \draw [->] (0,0) -- (20,0) node[align=center, below] {time $j$};
    \draw[dotted] (0,5) node[left] {$\sum\limits_{j=1}^{t_i} \frac{c_i(j)}{t_i}$} -- (20,5);
    \draw (0,10) node[left] {$h_i$} -- (20,10);
    \draw (20,7.5) node[right] {\scalebox{1.1}[3.5]{\}}};
    \draw (21,7.5) node[right] {~$C$};
    \draw (0,7.5) parabola [bend at end] (1,8) parabola (2,7.5) parabola [bend at end] (3,7) parabola (4,8) parabola [bend at end] (5,9) parabola (6,8) parabola [bend at end] (7,7) parabola (8,6.5) parabola  [bend at end] (9,6) parabola (10,5.5) parabola [bend at end] (11,5) parabola (12,5.5) parabola  [bend at end] (13,6) parabola (14,5) parabola  [bend at end] (15,4) parabola (16,3.5) parabola [bend at end] (17,3) parabola (18,2.5) parabola [bend at end] (19,2) parabola (20,3);
\end{tikzpicture}
    \caption{Relation of hint $h_i$ to average information gain $\sum_{j=1}^{t_i} c_i(j)/t_i$ as in Def.~\ref{def:accuracy}.}
    \label{fig:relation}
\end{figure}

We measure the accuracy of hints with parameter $C$ from Def.~\ref{def:accuracy}. This relation is illustrated in Fig.~\ref{fig:relation}. 
Lem.~\ref{lemma:lowerbound_c_upper_cbar} provides a lower bound in dependence on this accuracy.

\begin{definition}[Accuracy of Hints] \label{def:accuracy}
The accuracy $C\geq 1$ is the smallest number s.t.
\[
h_i \leq C \cdot \sum_{j=1}^{t_i} \frac{c_i(j)}{t_i} \quad\quad \forall i\in[n] \enspace.
\]
\end{definition}

The hint is a single number giving a cue about a function. Therefore, no matter which measure of accuracy we consider, if the hint is perfectly accurate, the function has to be constant. In Section~\ref{sect:conclusion}, we discuss other ideas on the measure of accuracy of hints such as a multi-dimensional feature vector or the hint being a random variable drawn from the information rate.

We introduce an auxiliary problem to bound the algorithm's expected value. 
\begin{definition}[Knapsack Problem for Hints (\kph{})]
Given an instance $I$ for \rao{}, i.e., time budget $T$, hint $h_i$ and length $t_i$ for each article $i\in [n]$,
the Knapsack Problem for Hints, \kph{}, is the fractional knapsack problem
with values $t_ih_i$, weights $t_i$ and knapsack size $T$. 
Let $\Opt_{\kph{}}(I)$ denote its optimal value on
instance $I$.
\end{definition}

As in \cite{KnapsackSecretaryProblem}, we now define an LP for a given subset $Q\subseteq [n]$
and time budget $x$. It finds the optimal fractional
solution for \kph{} with articles from $Q$ and time budget $x$.
\begin{equation*}
\begin{array}{rrrcll}
 \max & \sum_{i=1}^n & t_i\cdot h_{i}\cdot y(i)\\[.2cm] 
\textrm{s.t.} & \sum_{i=1}^n &t_i\cdot  y(i) & \leq & x \\
& & y(i) &= &0 &\quad\forall i\notin Q \\
& & y(i) &\in & [0,1] &\quad\forall i\in [n]
\end{array}
\end{equation*}

The variable $y_Q^{(x)}(i)$ refers to the setting of $y(i)$ in the optimal solution on articles from  $Q$ with time budget $x$.
The optimal solution has a clear structure: There exists a threshold hint $\rho_Q^{(x)}$ such that
any article $i\in Q$ with $h_i>\rho_Q^{(x)}$ has $y_Q^{(x)}(i)=1$ and any article
$i\in Q$ with $h_i<\rho_Q^{(x)}$ has $y_Q^{(x)}(i)=0$.
As in \cite{KnapsackSecretaryProblem}, we use the following notation for a subset  $R\subseteq [n]$:

\[
v_Q^{(x)}(R) = \sum_{i\in R}t_i\cdot  h_i\cdot  y_Q^{(x)}(i) \quad\text{ and }\quad
w_Q^{(x)}(R) = \sum_{i\in R} t_i \cdot  y_Q^{(x)}(i) \enspace.
\]

The value $v_Q^{(x)}(R)$ and weight $w_Q^{(x)}(R)$ refer to the value and weight that set $R$ contributes to the optimal solution.
We use \kph{}'s solution as an upper bound on the optimal solution, as shown in the following lemma:

\begin{lemma}\label{lemma:relationOpts}
Given instance $I$ for \rao{}, let $\Opt(I)$ and $\Opt_{\kph{}}(I)$ be 
the respective optimal values.
Then,
$\Opt_{\kph{}}(I)\geq \Opt(I).$
\end{lemma}
\begin{proof}
Since the codomain of $c_i$ is $[h_i]$ by Def.~\ref{def:RA}, any algorithm for \rao{} cannot
obtain more information units than $h_i$ in any time step.
Thus, the optimal solution of \kph{} obtains at least the same amount of information units as the optimal solution of \rao{} by reading the same parts. \qed
\end{proof}

\section{Lower Bounds} \label{sect:LBs}

The proofs in this section are constructed in a way
such that the instances are not producing a lower bound for the other settings.
Note that in the proofs of Lem.~\ref{lemma:increasing_functions} and Lem.~\ref{lemma:length}, we have $C\leq 2$.
Moreover, we construct the family of instances such that the lower bounds hold in a fractional setting. The key idea is to have the first time step(s)
small in every information rate such that the algorithms are forced to spend a minimal amount of time on reading an article before obtaining eventually more than one information unit per time step. 

Although Def.~\ref{def:RA} already states that the codomain of $c_i$ is $[h_i]$, 
we show a lower bound as a justification for this constraint on $c_i$.

\begin{lemma} \label{lemma:increasing_functions}
If the functions $c_i$ are allowed
to take values larger than hint $h_i$,
then the competitive ratio of any deterministic or randomized algorithm is $\Omega(\sqrt{n})$.
\end{lemma}

\begin{proof}
We construct a family of instances for all $\ell\in \nat$. Let $n:=\ell^2$. For any fixed $\ell$, we construct instance $I$ as follows.
Set $T:=n$, $t_i:=T=n$ and $h_i:=n$ for all $i\in[n]$.
We define two types of articles.
There are $\sqrt{n}$ articles of type $A$ where
$c_i(j):=1$ for $j\in[T]\setminus \set{\sqrt{n}}$ and $c_i(\sqrt{n}):=n^2$.
The articles of type $B$ have
$c_i(j):=1$ for $j\in[T-1]$ and $c_i(T):=n^2$.

An optimal offline algorithm reads all articles of type $A$ up to time step $\sqrt{n}$. Therefore, $\Opt(I) = \Theta(n^{2.5}).$
Any online algorithm cannot distinguish between articles of type $A$ and $B$ until time step $\sqrt{n}$.
The value of any online algorithm cannot be better than the value of algorithm $\ALG$ that reads the first $\sqrt{n}$ time
steps of the first $\sqrt{n}$ articles. Since the input order is chosen uniformly at random, the expected arrival of the first type $A$ article is at round $\sqrt{n}$.
Thus, $\expected{}{\ALG(I)} = \Theta(n^2)$. Therefore, $\expected{}{\ALG(I)} \leq \Opt(I)/\sqrt{n}.$ \qed
\end{proof}

In \rao{}, solutions admit reading articles fractionally.
Therefore, we show a lower bound whenever
articles are longer than the time budget.

\begin{lemma} \label{lemma:length}
If the lengths $t_i$ are allowed to take values larger than time budget $T$,
then the competitive ratio of any deterministic or randomized algorithm is $\Omega(n)$.
\end{lemma}

\begin{proof}
We construct a family of instances with $T:=2$, $t_i:=3$ and $h_i:=M$ for all $i\in [n]$, where $M\ge 1$ is set later. 
We define $c_k(1):=1$, $c_k(2):=M$ and $c_k(3):=1$.
Any other article $i\in [n]\setminus\set{k}$ has $c_i(1):=1$, $c_i(2):=1$ and $c_i(3):=M$.

As the permutation is chosen uniformly at random, 
an online algorithm does not know which article is the one with $M$ information units in the second time step.
No algorithm can do better than the algorithm $\ALG$ that reads the first article completely while $\Opt(I)=M+1$. 
Its expected value is
$\expected{}{\ALG(I)} = (1/n)\cdot (M+1) + (1-1/n)\cdot2 \leq (2/n + 2/M)\cdot \Opt(I).$
When setting $M=n$, we obtain the desired bound. \qed
\end{proof}

A consequence of the next lemma is that if the accuracy $C$ from Def.~\ref{def:accuracy} is not a constant, then no constant-competitive algorithms can exist.

\begin{lemma} \label{lemma:lowerbound_c_upper_cbar}
Any deterministic or randomized algorithm is $\Omega(\min\set{C,n})$-com\-pe\-ti\-ti\-ve. 
\end{lemma}

\begin{proof}
Consider the following family of instances $I$ in dependence on accuracy $C\geq 1$.
Let $T:=2$ and $t_i:=2$ for all $i\in[n]$.
Set $c_k(1):=1$ and $c_k(2):=C$. We define $c_i(1):=1$ and $c_i(2):=1$ for all $i\in[n]\setminus\set{k}$.
The hints are $h_i:=C$ for all $i\in[n]$, thus, they are $C$-accurate according to Def.~\ref{def:accuracy}.

Any algorithm cannot distinguish the information rate of the articles,
as the hints and the first time steps are all equal. Therefore, no algorithm is better 
than $\ALG$, which chooses to read the article arriving first completely.
The optimal choice is to read article $k$; we obtain the desired bound:
$\expected{}{\ALG(I)} = (1/n)\cdot (C+1) + (1-1/n)\cdot2 \leq (2/n + 2/C)\cdot \Opt(I).
$\qed
\end{proof}

\begin{assumption} \label{assum:Caverage}
For any article $i\in [n]$, we assume that $t_i\leq T$ and that the hints $h_i$ and upper bounds $t_ih_i$ on the information gain in the articles
are distinct.\footnote{Disjointness is obtained by random, consistent tie-breaking as described in \cite{KnapsackSecretaryProblem}.}
\end{assumption}

\section{Exploitation of Online Knapsack Algorithms}
In this section, we develop a technique for applying any algorithm for the Online Knapsack Problem on an instance of \rao{}.
The presented algorithm uses the classic Secretary Algorithm that is $\eee$-competitive for all positive $n$ as shown in \cite{MatroidSecretary}.
The Secretary Algorithm rejects the first $\lfloor n/\eee\rfloor$ items. 
Then it selects the first item that has a better value than the best so far. Note that the Secretary Algorithm selects exactly one item.

We use \kph{} for the analysis as an upper bound on the actual optimal solution with respect to information rates $c_i$. There is exactly one fractional item in the optimal solution of \kph{}. The idea is to make the algorithm robust against two types of instances: the ones with a fractional article of high information amount and the ones with many articles in the optimal solution.

\begin{theorem}\label{thm:reduction}
Given an $\alpha$-competitive algorithm $\ALG$ for the Online Knapsack Problem,
the Reduction Algorithm
is $(\eee+\alpha)C$-competitive.
\end{theorem}

\begin{proof}
We fix an instance $I$ and use Lem.~\ref{lemma:relationOpts}.
We split the optimal solution of \kph{} into the fractional article $i_{f}$ that is read $x_{i_{f}}\cdot t_{i_{f}}$ time steps and the set $H_{max}$ of articles
that are read completely.
Since $H_{max}$ is a feasible solution to the integral version of \kph{}, the value of the articles in $H_{max}$ is not larger than
the optimal value $\Opt_{\kph{}}^{int}(I)$ of the integral version. We denote the optimal integral solution by set $H^*$.
Using Def.~\ref{def:accuracy}, we obtain:

\begin{equation*} \label{eq:reduction}
\begin{aligned}
\Opt(I) &\leq \Opt_{\kph{}}(I) = \sum_{i\in H_{max}} t_{i}h_i + x_{i_{f}}t_{i_{f}}h_{i_{f}}
\leq \Opt_{\kph{}}^{int}(I) +  t_{i_{f}}h_{i_{f}}\\
&\leq \sum_{i\in H^*} t_{i}h_i + \max_{i\in[n]} t_{i}h_i
\leq C\cdot \left(\sum_{i\in H^*} \sum_{j=1}^{t_i} c_i(j) + \max_{i\in[n]} \sum_{j=1}^{t_i}c_i(j)\right)\\
&\leq C\cdot \left(\frac{\alpha}{\delta}\cdot  \prob{}{b=1}\expected{}{\sum_{i\in S} \sum_{j=1}^{t_i}c_i(j)\bigg|b=1}
\right. 
\\
& \quad\quad\quad\quad \left.
+ \frac{\eee}{1-\delta} \cdot \prob{}{b=0}\expected{}{\sum_{i\in S} \sum_{j=1}^{t_i}c_i(j)\bigg|b=0}\right)\\
&\leq C\cdot \max\set{\frac{\alpha}{\delta},\frac{\eee}{1-\delta}}\cdot \expected{}{\text{Reduction Algorithm}(I)}\enspace.
\end{aligned}
\end{equation*}

The optimal choice of $\delta$ to minimize $\max\set{\frac{\alpha}{\delta},\frac{\eee}{1-\delta}}$
is $\delta = \frac{\alpha}{\eee+\alpha}$.
This is exactly how the Reduction Algorithm sets the probability $\delta \in (0,1)$ in line 1, which yields a competitive ratio of $(\eee+\alpha)\cdot C$.\qed
\end{proof}

\begin{algorithm}
\label{algo:reduction}
\caption{Reduction Algorithm}
\KwIn{Number of articles $n$, time budget $T$, an $\alpha$-competitive algorithm $\ALG$ for the Online Knapsack Problem.}
\KwOut{Set $S$ of chosen articles.\medskip}
Set $\delta=\frac{\alpha}{\eee+\alpha}$ and choose $b\in \set{0,1}$ randomly with $\prob{}{b=1}=\delta$\;
\eIf{$b=1$}{
Apply $\ALG$ with respect to values $t_ih_i$ and weights $t_i$\;
}
{Apply the Secretary Algorithm with respect to values $t_ih_i$\;}
\end{algorithm}

When using the current best algorithm for the Online Knapsack Problem presented in \cite{AlbersKnapsackGAP}, the Reduction Algorithm has a competitive ratio of $(\eee+6.65)\cdot C\leq 3.45\eee C$.
Assuming that the accuracy of the hints $C\geq 1$ from Def.~\ref{def:accuracy} is constant, the \rao{} admits a constant upper bound on the competitive ratio.

\begin{remark}
(i) The Reduction Algorithm can be used to obtain an $(\alpha+\eee)$-com\-pe\-ti\-ti\-ve algorithm for the
fractional version of the Online Knapsack Problem given an $\alpha$-competitive algorithm 
for the integral version.
The proof is analogous to the proof of Thm.~\ref{thm:reduction}.
(ii) Running an $\alpha$-competitive algorithm for Online Knapsack on 
an instance of \rao{}, we obtain a $2\alpha C$-competitive algorithm for \rao{} by a similar proof. Since the current best algorithm for Online Knapsack has $\alpha=6.65>\eee$, 
using the Reduction Algorithm provides better bounds.
This holds for the fractional Online Knapsack Problem respectively.
\end{remark}

\section{Threshold Algorithm} \label{sect:threshold}
While the Online Knapsack Problem has to take items completely, \rao{}
does not require the reader to finish an article.
Exploiting this possibility, we present 
the Threshold Algorithm, which bases its
decisions on a single threshold. 
We adjust the algorithm and its analysis from \cite{KnapsackSecretaryProblem}.
From now on, we assume that the information rates $c_i$ are non-increasing and that we can stop to read an article at any time, thus allowing fractional time steps. For the sake of presentation, we stick to the discrete notation (avoiding integrals). 
In practice,
the information gain diminishes the longer an article is read, and the inherent discretization by words or characters is so fine-grained compared to the lengths of the articles that it is justified to consider the continuum limit.

Before starting to read, one has to decide at which length to stop reading any article in dependence on $T$.
First, we show that cutting all articles of an instance after $gT$ time steps costs at most a factor of 
$1/g$ in the competitive ratio.

\begin{lemma}\label{lemma:cutInstance}
Given an instance $I$ with time budget $T$, $g\in(0,1]$, lengths $t_i$, hints $h_i$ and non-increasing information rates $c_i:[t_i]\to[h_i]$,
we define the cut instance $I'_g$ with
time budget $T'=T$, lengths $t'_i=\min\set{t_i,gT}$, hints $h'_i=h_i$ and non-increasing information rates $c'_i:[t'_i]\to[h'_i]$,
where $c'_i(j)=c_i(j)$ for $1\leq j\leq t'_i$.
Then, $\Opt_{\kph{}}(I)\leq \Opt_{\kph{}}(I'_g)/g$.
\end{lemma}

\begin{proof}
Since $gt_i\leq gT$ and $gt_i\leq t_i$ we have $gt_i\leq \min\set{gT,t_i}=t'_i$ and obtain:

\noindent $
\Opt_{\kph{}}(I)
= \frac{1}{g}\sum_{i\in[n]} h_i \cdot gt_i  \cdot y_{[n]}^{(T)}(i)
\leq \frac{1}{g}\sum_{i\in[n]} h'_i t'_i \cdot y_{[n]}^{(T)}(i)
\leq \frac{1}{g}\cdot \Opt_{\kph{}}(I'_g).
$
The last inequality follows as no feasible solution is better than the optimum.
The time budget is respected since $\sum_{i\in[n]} t_i \cdot y_{[n]}^{(T)}(i)\leq T=T'$ and $t_i\geq t'_i$.\qed
\end{proof}

\begin{algorithm} \label{algo:threshold}
\caption{Threshold Algorithm}
\KwIn{Number of articles $n$, time budget $T$, a fraction $g\in (0,1]$.\\
\hspace*{1.05cm} Article $i$ appears in round $\pi(i)$ and reveals $h_i$ and $t_i$.}
\KwOut{
Number of time steps $0\leq s_i\leq t_i$ that are read from article $i\in [n]$.\medskip}
Sample $r\in \set{1,...,n}$ from binomial distribution $\Bin(n,1/2)$\;
Let $X=\set{1,...,r}$ and $Y=\set{r+1,...,n}$\;
\For{round $\pi(i)\in X$}{Observe $h_i$ and $t_i$\;
Set $s_i=0$\;}
Solve \kph{} on $X$ with budget $T/2$, lengths $t'_i=\min\set{gT,t_i}$, and values $t'_ih_i$. \\
Let $\rho_X^{(T/2)}$ be the threshold hint\;
\For{round $\pi(i)\in Y$}{
\uIf{ $h_i\geq \rho_X^{(T/2)}$}{
Set $s_i =\min\left\{t_i,gT, T-\displaystyle\sum_{1\leq j<i} s_j\right\}$\;
Read the first $s_i$ time steps of article $i$\;}
\uElse{Set $s_i=0$\;}
}
\end{algorithm}

We need the following lemma that can be proven by a combination of normalization, Exercise~4.7 on page~84 and Exercise~4.19 on page~87 in \cite{ProbabilityAndComputing}.

\begin{lemma}\label{lemma:chernoff}
Let $z_1, ..., z_n$ be mutually independent random variables from a finite subset of $[0,z_{max}]$ and $Z=\sum_{i=1}^n z_i$ with $\mu = \expected{}{Z}$.
For all $\mu_H\geq \mu$ and all $\delta>0$,
\[\prob{}{Z\geq (1+\delta)\mu_H}< \exp\left(  -\frac{\mu_H}{z_{max}}\cdot
\left[ (1+\delta) \ln(1+\delta) -\delta \right]\right)\enspace.\]
\end{lemma}

We assume that 
$\sum_{i\in[n]} \min\set{t_i, gT} = \sum_{i\in[n]} t'_i \ge 3T/2$ for the purpose of the analysis. The same assumption is made in \cite{KnapsackSecretaryProblem} implicitly.
If there are not enough articles, the algorithm can only improve since there are fewer articles that are not part of the optimal solution.
We can now state the main theorem.

\begin{theorem}\label{theorem:threshold}
For $g = 0.0215$, the Threshold Algorithm's competitive ratio is upper bounded by $ 246 C<90.5 \eee C$.
\end{theorem}

The proof is similar to the proof of Lem.~4 in \cite{KnapsackSecretaryProblem}.
However, we introduce parameters over which we optimize the analysis of the competitive ratio.
This way, we make the upper bound on the competitive ratio as tight as possible for the proof technique used here.
Recall that we may assume w.l.o.g.~by Assumption~\ref{assum:Caverage} that the hints $h_i$ and upper bounds $t_ih_i$ are disjoint throughout the proof.

\begin{proof}
Fix an instance $I$.
We refer to the order by permutation $\pi$ chosen uniformly at random.
For simplicity, we scale the instance such that $T=1$ and all $t_i$ are multiplied with $1/T$, which does not affect the hints and the threshold.
We use \kph{} to show the bound
as $\Opt_{\kph{}}(I)\geq \Opt(I)$ holds by Lem.~\ref{lemma:relationOpts}.
As the reader always reads at most the first $gT$ time steps,
we use the bound from Lem.~\ref{lemma:cutInstance} for cutting instance $I$ to $I'_g$.
For better readability, we do not rename the parameters of $I'_g$ and refer to the variables without adding a prime $'$.
We proceed with showing the bound on the expected value of the algorithm on instance $I'_g$ since the algorithm reads at most $g$ time steps of each article.

Now, we proceed as in the proof of Lem.~4 in \cite{KnapsackSecretaryProblem}.
We use two auxiliary knapsacks to bound the algorithm's expected value.
Their optimal, fractional solution is computed offline on instance $I$.
In contrast to \cite{KnapsackSecretaryProblem}, we parameterize the size of the auxiliary knapsacks to find the best possible sizes.
We use a knapsack of size $\beta$ and one of size $\gamma$ where $0<\beta\leq 1$ and $1\leq \gamma\leq \sum_{i\in[n]} t_i$.
Recall that we assumed that $\sum_{i\in[n]} t_i \ge 3/2 = 3T/2$.
We show in the following that for all $i$ where $y_{[n]}^{(\beta)}(i)>0$, there is a $p\in (0,1)$ such that $\prob{}{s_i = t_i}>p$. As a consequence of $p$'s existence, we obtain the following inequalities: 
\begin{equation}\label{eq:competitive}
\begin{aligned}
\Opt_{\kph{}}(I) &\leq \frac{1}{g} \cdot \Opt_{\kph{}}(I'_g)
\leq \frac{1}{g} \cdot \frac{1}{\beta} \cdot v_{[n]}^{(\beta)}([n])= \frac{1}{g\beta} \cdot \sum_{i\in [n]} t_ih_i y_{[n]}^{(\beta)}(i)\\
&\leq \frac{1}{g\beta p} \cdot \sum_{i\in [n]} \prob{}{s_i= t_i} t_ih_i
\leq \frac{1}{g\beta p} \cdot \sum_{i\in [n]} \prob{}{s_i= t_i} \cdot C \cdot \sum_{j=1}^{s_i}c_i(j)\\ 
&\leq \frac{C}{g\beta p} \cdot \expected{}{\text{Threshold Algorithm}(I)}\enspace.\\
\end{aligned}
\end{equation}

We lose the factor $C$ as we use the inequality from Def.~\ref{def:accuracy}.
The best possible value for the competitive ratio is the minimum of $C/(g\beta p)$.
We find it by maximizing $g\beta p$.
As $g$ and $\beta$ are settable variables, we determine $p$ first. 
We define random variables $\zeta_i$ for all $i\in[n]$, where
\[\zeta_i = \left\{ \begin{array}{ll}
1  & \mbox{if } \pi(i)\in X\\
0 & \mbox{otherwise}\enspace.
\end{array}
\right.\]
As discussed in \cite{KnapsackSecretaryProblem}, 
conditioned on the value of $r$, $\pi^{-1}(X)$ is a uniformly chosen subset of $[n]$ from any subset of $[n]$ containing exactly $r$ articles.
Since $r$ is chosen from $\Bin(n,1/2)$, it has the same distribution as the size of a uniformly at random chosen subset of $[n]$.
Therefore, $\pi^{-1}(X)$ is a uniformly chosen subset of all subsets of $[n]$.
The variables $\zeta_i$ are mutually independent Bernoulli random variables with $\prob{}{\zeta_i=1}=1/2$.
Now, we fix $j\in[n]$ with $y_{[n]}^{(\beta)}(j)>0$ and define two random variables: 
\begin{align*}
Z_1 &:= w_{\pi([n])}^{(\beta)}(X\setminus \set{\pi(j)})
= \sum_{i\in [n]\setminus\set{j}} t_i\cdot y_{[n]}^{(\beta)}(i)\cdot\zeta_i\enspace,\\
Z_2 &:= w_{\pi([n])}^{(\gamma)}(Y\setminus \set{\pi(j)})
= \sum_{i\in {[n]}\setminus\set{j}} t_i\cdot y_{[n]}^{(\gamma)}(i)\cdot(1-\zeta_i) \enspace.
\end{align*}

Note that the event $\pi(j)\in Y$, i.e., $\zeta_j=0$, is independent of $Z_1$ and $Z_2$ since they are defined without $\pi(j)$.
The weights $t_i\cdot y_{[n]}^{(\beta)}(i)\cdot\zeta_i$ and $ t_i\cdot y_{[n]}^{(\gamma)}(i)\cdot(1-\zeta_i)$ within the sum are random variables taking values in $[0,g]$ since the instance is cut.
Now, we reason that
when article $j$ is revealed at position $\pi(j)$ to the Threshold Algorithm, 
it has enough time to read $j$ with positive probability.
The next claim is only effective for $g<0.5$ since $Z_1$ and $Z_2$ are non-negative.

\begin{claim}\label{claim:conditionalEta}
Conditioned on  $Z_1<\frac{1}{2}-g$ and $Z_2< 1-2g$, 
the Threshold Algorithm sets $s_j=t_j$ 
if $\pi(j)\in Y$ because $h_j\geq \rho_X^{(1/2)}$ and it has enough time left.
\end{claim}

\begin{proof}
Since $t_j\leq gT=g$, article $j$ can only add at most $g$ weight to $X$ or $Y$.
Therefore, $w_{\pi([n])}^{(\beta)}(X)<1/2$ and $w_{\pi([n])}^{(\gamma)}(Y)<1-g$.
Recall that knapsacks of size $\beta$ and $\gamma$ are packed optimally and fractionally. Thus, a knapsack of size $\gamma$ would be full, i.e.,
$w_{\pi([n])}^{(\gamma)}(\pi([n]))=\min\set{\gamma, \sum_{i\in [n]} t_i}$.
Since we assumed in the beginning of the proof that $\gamma\leq \sum_{i\in [n]} t_i$, we have $w_{\pi([n])}^{(\gamma)}(\pi([n]))=\gamma$. 
We can bound the weight of $X$ in this solution by $w_{\pi([n])}^{(\gamma)}(X) = \gamma- w_{\pi([n])}^{(\gamma)}(Y)>\gamma-(1-g)$. We choose $\gamma$ such that
$\gamma-(1-g) >1/2.$\footnote{In \cite{KnapsackSecretaryProblem}, it is implicitly assumed that $\sum_{i\in [n]} t_i\ge 3/2$ as their choice of $\gamma$ is $3/2$.}
Thus, the articles in $X$ add weights $w_{\pi([n])}^{(\beta)}(X) <1/2$ and $w_{\pi([n])}^{(\gamma)}(X) >1/2$ to their respective optimal solution. 
Since $w_{\pi([n])}^{(\gamma)}(X) = \sum_{i\in X} t_iy^{(\gamma)}_{n}(i) >1/2$, it means that there are elements in $X$ with combined $t_i$ of at least $1/2$. Therefore, the optimal solution of \kph{} on $X$ with time budget $1/2$ is satisfying the capacity constraint with equality, i.e.,  $w_{X}^{(1/2)}(X) =1/2$.
We obtain:
$w_{\pi([n])}^{(\gamma)}(X)>w_{X}^{(1/2)}(X)>w_{\pi([n])}^{(\beta)}(X).$

When knapsack~$A$ has a higher capacity than knapsack~$B$ on the same instance, 
then the threshold density of knapsack~$A$ cannot be higher than the threshold density of knapsack~$B$.
For any $X$, we see that the knapsack of size $\beta$ uses less capacity than $1/2$ and the knapsack of size $\gamma$ uses more capacity than $1/2$ on the same $X$ respectively.
Since both knapsacks are packed by an optimal, fractional solution
that is computed offline on the whole instance, the respective threshold hint for articles from $X$ and articles from $\pi([n])$ is the same.
Therefore, we get the following ordering of the threshold hints:
$\rho_{\pi([n])}^{(\gamma)}\leq \rho_{X}^{(1/2)} \leq \rho_{\pi([n])}^{(\beta)}.$

Now, we show that when the algorithm sees $\pi(j)$, it has enough time left.
Let 
$S^+ = \set{\pi(i)\in Y\setminus \set{\pi(j)}: h_i\geq \rho_{X}^{(1/2)}}$ be the set of articles that the algorithm can choose from. 
Thus, the algorithm reads every article from $S^+$ (and maybe $\pi(j)$) if it has enough time left at the point when an article from $S^+$ arrives.
By transitivity, every article $\pi(i)\in S^+$ has $h_i\geq \rho_{\pi([n])}^{(\gamma)}$.
Therefore, for all but at most one\footnote{as the hints are distinct by Assumption~\ref{assum:Caverage}}
$\pi(i)\in S^+$, the equation $y_{[n]}^{(\gamma)}(i)=1$ holds.
Since the only article $i\in S^+$ that could have $y_{[n]}^{(\gamma)}(i)<1$ is not longer than $g$, 
the total length of articles in $S^+$ can be bounded from above by
$\sum_{i\in S^+} t_i \leq g + w_{\pi([n])}^{(\gamma)}(Y\setminus\set{\pi(j)}) = g+Z_2<1-g.$
As $t_j\leq g$, the algorithm has enough time left to read article $j$ completely, when it arrives at position $\pi(j)$.
Moreover, if $y_{[n]}^{(\beta)}(j)>0$, then $h_{j}\geq \rho_{\pi([n])}^{(\beta)}\geq \rho_{X}^{(1/2)}$. 
{\hfill$\blacksquare$}
\end{proof}

We now proceed with the main proof by showing a lower bound on 
$p'=\mathbb{P}[Z_1< 1/2-g \text{ and } Z_2< 1-2g]$. 
Recall that the event $\pi(j)\in Y$ is independent of $Z_1$ and $Z_2$.
With the preceded claim we obtain: 
\begin{equation}\label{eq:pprim} 
\begin{aligned}
 \prob{}{s_j=t_j} &= \prob{}{s_j=t_j\big| Z_1<1/2-g \text{ and } Z_2<1-2g}\cdot p'\\
 &= \prob{}{\pi(j)\in Y\big| Z_1<1/2-g \text{ and } Z_2<1-2g}\cdot p' =  \frac{1}{2}\cdot p'\enspace.
\end{aligned}
\end{equation}
As we are searching for the lower bound $p<\prob{}{s_j=t_j}$, we use the lower bound for $p'$ multiplied with $1/2$ as the value for $p$.
Moreover, 
\begin{equation}\label{eq:primbound}
\begin{aligned}
p' &=\prob{}{Z_1< 1/2-g \text{ and } Z_2< 1-2g} \\
&\geq 1-\prob{}{Z_1\geq 1/2-g} - \prob{}{ Z_2\geq 1-2g}.
\end{aligned}
\end{equation}

We can bound the probabilities $\prob{}{Z_1\geq 1/2-g}$ and $\prob{}{ Z_2\geq 1-2g}$ by the Chernoff Bound from Lem.~\ref{lemma:chernoff}.\footnote{Note that the
Chernoff Bound is indeed applicable since the random variables $t_iy_{[n]}^{(\beta)}(i)\zeta_i$ and $t_iy_{[n]}^{(\gamma)}(i)(1-\zeta_i)$ are discrete and $\zeta_i$ are mutually independent.}
The expected values of $Z_1,Z_2$ are bounded by
\[
\expected{}{Z_1} =
\frac{1}{2}\cdot\left(\beta - t_j\cdot y_{[n]}^{(\beta)}(j) \right) \leq \frac{\beta}{2} \quad \text{ and } \quad\quad \\
\expected{}{Z_2} = 
\frac{1}{2}\cdot\left(\gamma - t_j\cdot y_{[n]}^{(\gamma)}(j) \right) \leq \frac{\gamma}{2} \enspace.
\]

\noindent We use $z_{max}=g$, 
$\delta_1 = (1-2g)/\beta-
1>0 \text{ and } \delta_2 = (2-4g)/\gamma - 1>0$
to obtain:

\begin{equation}\label{eq:pr1pr2}
\begin{aligned}
\prob{}{Z_1\geq 1/2-g} 
&<\exp \left(\left(1-\frac{1}{2 g}\right) \cdot \ln \left(\frac{1-2 g}{\beta}\right)-1+\frac{1-\beta}{2 g}\right)\enspace,\\[.3cm]
\prob{}{ Z_2\geq 1-2g}
&<\exp \left(\left(2-\frac{1}{g}\right) \cdot \ln \left(\frac{2-4g}{\gamma}\right)-2+\frac{1-\gamma/2}{g}\right)\enspace.
\end{aligned}
\end{equation}

To conclude the proof, 
the final step is numerically maximizing the lower bound on $g\beta p'/2$
obtained by combining Equations~(\ref{eq:pprim}),~(\ref{eq:primbound})~and~(\ref{eq:pr1pr2}):
\begin{samepage}
\begin{equation}\label{eq:maximization}
\begin{array}{r@{\quad}l}
\displaystyle \max_{g,\beta,\gamma} &\frac{\beta g}{2} \cdot \bigg(1-\exp \Big(\big(1-\frac{1}{2 g}\big) \cdot \ln \big(\frac{1-2 g}{\beta}\big)-1
+\frac{1-\beta}{2 g}\Big) \\
& \quad\quad \quad\ - \exp \Big(\big(2-\frac{1}{g}\big) \cdot \ln \big(\frac{2-4g}{\gamma}\big)-2 +\frac{1-\gamma/2}{g}\Big) \bigg)
\end{array} \\
\end{equation} \vspace*{-0.25cm}
\begin{equation*}
\begin{array}{r@{\quad\quad}l@{\quad\quad}l@{\quad\quad}l}
\textrm{s.t.} & \gamma+g>1.5 & 2g+\beta<1 & 4g+\gamma<2 \\ 
& 0<g< 0.5 & 0<\beta < 1  & \gamma > 1
\end{array}
\end{equation*}
\end{samepage}
We do not use $\gamma\leq \sum_{i\in[n]} t_i$ as a constraint because the other constraints already imply that $\gamma < 2$, and we assume that the combined length of all articles is huge compared to the time budget.
Numerical maximization of (\ref{eq:maximization}) using \cite{maxima}
yields 
$g~=~0.021425,$ $\beta~=~0.565728,$ $ \gamma~=~1.478575.$
As we set $p$ to the lower bound on $p'/2$, we can plug
these values in Equation~(\ref{eq:competitive}), so
$
\Opt(I)< 246 C \cdot \expected{}{\text{Threshold Algorithm}(I)}.
$
{\qed}
\end{proof}

It is interesting to note that our proof suggests to limit the time for each article to about 2\% of the time budget in order to maximize the expected total information gain.
The analysis from Lem.~4 in \cite{KnapsackSecretaryProblem} uses $\beta=3/4$, $\gamma=3/2$ and $t_i \le 1/81$ for all $i\in[n]$.
Applying their analysis on the Threshold Algorithm for $g=1/81$ yields an upper bound of $162\eee C$ on its competitive ratio.
For these parameters, the optimized analysis in Thm.~\ref{theorem:threshold} gives an upper bound of $125.77\eee C$.

\section{Open Questions} \label{sect:conclusion}
An open question is whether the analysis of the Threshold Algorithm is tight.
Although we optimize to find the best possible $g$, the used Chernoff bound is not applicable for $g\geq 1/6$. Moreover, the combined articles' lengths, cut with respect to $g$, have to be at least $1.48$ times greater than the time budget.
For the sake of improving the analysis, a different approach has to be investigated.

We informally related diminishing information gain over time to efficient reading strategies. It would be interesting to formalize it w.r.t.~spatial information profiles. 
Further directions involve the exploration of new settings and extensions.
There are different measures of the accuracy of the hints worth investigating. 
An example would be to interpret the information rate as a distribution of information and the hint to be a random variable drawn from this distribution. 
Then, the algorithm's performance is dependent on the information rate's expectation and standard deviation.
An interesting task is to develop an algorithmic strategy for the setting where the length $t_i$ of an article
is not revealed when it arrives.
We believe that the studied techniques in this paper can be used for this setting
if the information gain diminishes, e.g., logarithmically as a function of time, while reading any article.
Another reasonable setting is the one where articles appear in a non-uniform, but still random order.
This is suitable for reading articles since many websites present articles in a categorized order or using recommender systems, where articles are sorted based on the user's preferences. In that light, it would also make sense to extend the scalar hint to a multi-dimensional feature vector. The investigation of related \emph{learning-augmented} online algorithms would be a further interesting development. The idea of a threshold can be considered in that direction: Instead of the learning phase in the Threshold Algorithm, an external threshold can be considered, e.g., from past experience, gut feeling, or rating by a recommender system.

In our opinion, the most interesting extension is to allow the reader to mark a restricted number of articles and return to these articles at any point in time. 
For secretary problems with submodular objective functions, 
the setting where an algorithm is allowed to remember items and select the output after seeing the whole instance
has recently been discussed by \cite{shortlist}. 
Here, they achieve a competitive ratio that is arbitrarily 
close to the offline version's lower bound on the approximation factor. 
This extension combined with a cross reading strategy, unknown reading lengths, and articles sorted by categories or preferences is the closest setting to real-life web surfing.

\bibliographystyle{splncs04}% the mandatory bibstyle
\bibliography{references}

\appendix

\end{document}